# Space-variant polarized Airy beam

Hao Chen[1] and Guoqiang Li[1, 2, *]

[1]*Department of Ophthalmology and Visual Science, The Ohio State Universit,y, Columbus, OH 43212, USA*
[2]*Department of Electrical and Computer Engineering, The Ohio State University, Columbus, OH 43212, USA*

[*]*Corresponding author*: *li.3090@osu.edu*

**Abstract:** We experimentally generate an Airy beam with polarization structure while keeping its original amplitude and phase profile intact. This class of Airy beam preserves the acceleration properties. By monitoring their initial polarization structure we have provided insight concerning the self-healing mechanism of Airy beams. We investigate both theoretically and experimentally the self-healing polarization properties of the space-variant polarized Airy beams. Amplitude as well as the polarization structure tends to reform during propagation in spite of the severe truncation of the beam by finite apertures.

# 1. Introduction

Siviloglou et al and Besieris et al [1-3] first reported that optical Airy beam can remain diffraction free while its main intensity lobe freely self-bends (or transversely accelerates) during propagation. This class of Airy beam is characteristic of shape-preserving light beams whose peak intensity follows a continuous parabolic curve as they propagate in free space. Moreover, the two intriguing properties (self-bending and non-diffracting) of the Airy beams have attracted much interest quite recently in applications such as optical clearing of microparticles [4-6], curved plasma channel generation, and generating versatile linear bullets [7-9]. However, most reported work on the Airy beam is only limited to linearly polarized state up to now. On the other hand, owing to the unique polarization distribution, the properties of vector beams with various polarization states in its cross-section have attracted much research interest [10]. Polarization propagation and interaction with materials have been extensively explored in optical inspection and metrology, display technologies, data storage, optical communications, and materials sciences, as well as in biological studies [11-14]. To our best knowledge, polarization property of optical Airy beams has never been addressed before, owing to the lack of an effective way for modulating the amplitude, phase and polarization at the same time. In this paper, a two-dimensional optical Airy beams with space variant polarization state is generated and the general propagation dynamics of the space variant polarized Airy beam is demonstrated. It is found that, this type of Airy beam keep the shape-preserving property and acceleration property and the ability to reconstruct itself during propagation even though part of the beam is distorted or obstructed.

# 2. Generation of space-variant polarized Airy beam

Airy beams have previously been generated using a cubic phase pattern that represents the Fourier transform of the Airy beam [1,2,15]. The Fourier transform of this pattern is formed using a system length of 2f, where f is the focal length of the Fourier transform lens. Another approach for generating Airy beam by encoding the amplitude and phase proportional to the Fourier spectra of the desired beams in an input plane is proposed by Davis and his coworkers [15]. In previous work, space-variant polarized Airy beams (SVPAB) were generated in a 4f system. The experimental arrangement is similar to the configuration in Ref. 16 (Fig. 1 there).

A collimated linearly polarized laser beam illuminates a computer-controlled SLM, then coming into a 4f system composed of a pair of identical lenses with the focal length of f. A two-dimensional (2D) holographic grating (HG) is displayed on the SLM. The HG can be conceived as a superposition of two one-dimensional gratings oriented along x and y directions, respectively, each carrying its respective phase distribution. Two selected orders at the Fourier plane are converted into the left- and right-hand circularly polarized beams and recombined at the rear plane of the 4f system. The specialized system enables the independent modulation of amplitude, polarization and phase of an optical beam. The resulting beam is expressed by

$$\vec{E}(x,y) = \begin{pmatrix} E_x(x,y) \\ E_y(x,y) \end{pmatrix} = Ai(x,y)\begin{pmatrix} \cos\alpha(x,y) \\ \sin\alpha(x,y) \end{pmatrix} = A(x,y)e^{i\beta(x,y)}\begin{pmatrix} \cos\alpha(x,y) \\ \sin\alpha(x,y) \end{pmatrix}, \quad (1)$$

where $Ai(x,y)$ is the Airy function and $\alpha(x,y)$ is the local polarization angle of each point $(x,y)$ in the output plane. It is known that Airy beam has a specialized amplitude and phase profile, which is enveloped by $Ai(x,y)$ ($Ai(x,y) = A(x,y)e^{i\beta(x,y)}$). Two typical examples with different polarization states are demonstrated in Figs. 1-3, and the corresponding amplitude, phase, and polarization profile are also shown. We begin with a special case of the Airy beam with a polarization variant direction of 45° with respect to the x direction. The resulting intensity distribution of the beams is recorded by a charge-coupled device (CCD) camera. To analyze the polarization property of the beam, a rotating linear polarizer (its transmission axis shown in the inset) is placed in front of the CCD to provide the intensity pattern of the corresponding polarization component.

## 3. Propagation property of space-variant polarized Airy beam (SVPAB)

Let us now analyze the propagation property of the space-variant polarized Airy beam by a Fresnel diffraction [17]:

$$\vec{E}(x,y,z) = \begin{pmatrix} E_x \\ E_y \end{pmatrix} = \frac{ik}{2\pi z}\iint \exp\left(\frac{ik}{2z}\right)\left((x-x_1)^2 + (y-y_1)^2\right)\vec{E}(x,y,z=0)dx_1 dy_1 \quad (2)$$

From this equation, one can readily deduce the propagation of the SVPAB, which is decomposed of the two orthogonal polarized components.

First, an Airy beam with a polarization variant in the 45° is illustrated in Fig. 1. The inset arrows are identified as the local polarization direction. Linearly polarized Airy beam is also demonstrated for comparison. It is found that the shape of the SVPA is diffraction-free in the propagation, similar to the linearly polarized Airy beam. But the polarization state is changed dramatically. The polarization state tends to be uniform from the side lobe to the main lobe. Meanwhile, the polarization state of the main lobe is changed in the 45° with respect to the x axis. Second example is a radially polarized Airy beam illustrated in Fig. 2. The inset arrows are identified as the local polarization direction. It is found that the shape of the SVPA is diffraction-free in the propagation, similar to the linearly polarized Airy beam. Figure 3 presents the intensity profile and polarization state located at z = 0 and z = 20cm plane. In the same manner, the polarization feature tends to disappear in the free space propagation. It may be caused by the transverse energy flow directed to the corner of the profile.

So, we studied the Ponyting vector of the propagating space-variant Airy beam. It is interesting to note that in Figs. 4 and 5, the Poynting vector is initially pointing in the negative sx or negative sy-direction on each Airy tail; the direction starts to turn partially towards the direction of the energy flow of the main Airy peak (45°); and in as the beam propagates further, the direction swings around even more towards 45°. The change in the Poynting vector can be used to explain in Ref. And the transverse energy flow can be expressed by [18]

$$\vec{S} = \frac{c}{4\pi}\langle \vec{E}^* \times \vec{B}\rangle = \frac{c}{8\pi}\left(\vec{E}\times\vec{B}^* + \vec{E}^*\times\vec{B}\right) = \frac{c}{8\pi}\left[i\omega\left(u\nabla_T u^* - u^*\nabla_T u\right) + 2\omega|u|^2 \hat{z}\right] \quad (3)$$

The z-term in the above equation is the energy flow in the z-direction which is just proportional to the linear momentum density in that direction. This is typically the main contributing component. The first term is what we are really interested in here as it contributes a non-zero sx- and sy-component and an additional z-term to the Poynting vector. According to the Poynting vector profile presented in Fig. 4, it is found that the direction of the energy flow turns gradually towards the main Airy peak. The radially polarized Airy beam has the orthogonal polarized components in the two arms at the z = 0 plane. Overlap of the two orthogonal polarized components appears in the main lobe of the Airy beam after propagation of a certain distance (e.g., 20cm). As a result, the polarization state of the main lobe is polarized at 45° and at both sides of the main lobe the polarization state is tilted to the x and y axis and the intensity profile passing through the 135° analyzer demonstrates two weak side lobe besides the Airy peak.

The Airy beam is characteristic of self-healing property during the propagation in free space. SVPA has a self-healing ability during propagation. The most prominent intensity characteristic of an Airy beam happens to be its main corner lobe (as seen in Fig. 2) which contains a large percentage of the beam's total power. As we can see, the truncated ideal Airy beam keeps the shape and intensity of the ideal one (i.e. not truncated) until a certain distance, when it suddenly loses its shape. This suggests that this kind of apodization (finite aperture) is more effective than those made with exponential and Gaussian functions, where the resulting beams suffer intensity decay from the beginning.

## 4．Conclusions

Propagation of space-variant polarized Airy beams is studied through numerical calculations and optical experiments. The optical experiments demonstrate that the space-variant polarized Airy beam can preserve its original profile within a long distance range in free space propagation. In this work we have also demonstrated both theoretically and experimentally the self-healing properties of optical Airy beams. By studying their internal transverse energy flow we have provided an insight concerning the self-healing mechanism of the space-variant polarized Airy beams. We have also theoretically and experimentally studied the polarization change during the free space propagation. Our observations are in excellent agreement with the numerical simulations.

**References:**

1. G. A. Siviloglou, and D. N. Christodoulides, Accelerating finite energy Airy beams, Opt. Lett. 32, 979-981 (2007).
2. I. M. Besieris, and A. M. Shaarawi, A note on an accelerating finite energy Airy beam, Opt. Lett. 32, 2447-2449 (2007).
3. G. A. Siviloglou, J. Broky, A. Dogariu, and D. N. Christodoulides, Observation of Accelerating Airy Beams , Phys. Rev. Lett. 99, 213901 (2007).
4. J. Baumgartl, M. Mazilu and K. Dholakia, "Optically mediated particle clearing using Airy wavepackets," Nature Photonics 2, 675-678 (2008).
5. I. Chremmos, P. Zhang, J. Prakash, N. K. Efremidis, D. N. Christodoulides, and Z. Chen, Fourier-space generation of abruptly autofocusing beams and optical bottle beams , Optics Letters, Vol. 36, Issue 18, pp. 3675-3677 (2011)
6. J. Baumgartl, G. Hannappe, D. Stevenson, D. Day, M. Gu and K. Dholakia, "Optical redistribution of microparticles and cells between microwells," Lab Chip 9(2009) 1334-1336.
7. P. Polynkin, M. Kolesik, J.V. Moloney, G. Siviloglou and D. Christodoulides, "Curved Plasma Channel Generation Using Ultraintense Airy Beams," Sicence 324, 229-232 (2009).
8. P. Polynkin, M. Kolesik and J. Moloney, "Filamentation of Femtosecond Laser Airy Beams in Water," Phy. Rev. Lett. 103, 123902 (2009).
9. A. Chong, W. H. Renninger, D. Christodoulides and F. Wise, "Airy-Bessel wave packets as versatile linear light bullets," Nature Photonics 4, 103-106(2010).
10. H. Chen and G. Li, "Generation and application of vector beam with space-variant distribution of amplitude, polarization, and phase", The International Conference on Advanced Laser Applications in Science and Engineering, Nanjing, China, Nov. 1-3, 2012.
11. Q. Zhan, Adv. Opt. Photon. **1**, 1-57 (2009); and references therein.
12. S. Tripathi and K. C. Toussaint, "Rapid Mueller matrix polarimetry based on parallelized polarization state generation and detection," Opt. Express **17**, 21396 (2009).
13. M. Meier, V. Romano, and T. Feurer, "Material processing with pulsed radially and azimuthally polarized laser radiation," Appl. Phys., A Mater. Sci. Process. **86**, 329 (2007).
14. H. Chen, Z. Zheng, B. Zhang, J. Ding and H. Wang, "Polarization structuring of focused field

through polarization-only modulation of incident beam," Opt. Lett. **35**, 2825 (2010).

15. J. A. Davis, M. J. Mitry, M. A. Bandres, I. Ruiz, K. P. McAuley, and D. M. Cottrell, "Generation of accelerating Airy and accelerating parabolic beams using phase-only patterns," Appl. Opt. 48, 3170-3176 (2009).

16. H. Chen, J. Hao, B. Zhang, J. Xu, J. Ding, and H. Wang, "Generation of vector beam with space-variant distribution of both polarization and phase", Opt. Lett. **36**, 3179 (2011).

17. Born M and Wolf E, 1970, Principle of Optics (Oxford:Pergamon).

18. J. Broky, G. A. Siviloglou, A. Dogariu, and D. N. Christodoulides, Self-healing properties of optical Airy beams , Opt. Express 16, 12880-12891 (2008).

**Figure captions**

Fig.1. Generation functions $A(x,y), \alpha(x,y)$ and $\beta(x,y)$ are shown in (a), (b), and (c) respectively for an Airy beam with a polarization variant in 45° respect to the x axis. Corresponding intensity distribution at z = 0 (d - h) and z = 20cm plane (i - m) by passing a rotating analyzer.

Fig.2. Generation functions $A(x,y), \alpha(x,y)$ and $\beta(x,y)$ are shown in (a), (b), and (c) for an radially polarized Airy beam. Corresponding intensity distribution at z = 0 (d - h) and z = 20cm plane (i - m) by passing a rotating analyzer. Self-healing of a radially polarized Airy beam is demonstrated when its main lobe is blocked. Observed intensity profile at the input (n) z = 0, (o) z = 10cm, and (p) z = 20cm.

Fig.3. Linearly polarized Airy beam at (a - e) z = 0 and (f - j) z = 20cm plane, passing through a rotating analyzer.

Fig.4. Simulation results of linearly polarized Airy beam and radially polarized Airy beam at z = 0 plane (first column) and z = 20cm plane (second column). Corresponding energy flow in the z = 20cm plane is demonstrated. Yellow arrows indicate the local polarization state and green arrows indicate the local Poynting vector.

Fig.5. Simulation results of the blocked linearly polarized Airy beam and radially polarized Airy beam at z = 0 plane (first row) and z = 20cm plane (second column). Corresponding energy flow in z = 20cm plane is demonstrated. Yellow arrows indicate the local polarization state and green arrows indicate the local Poynting vector.

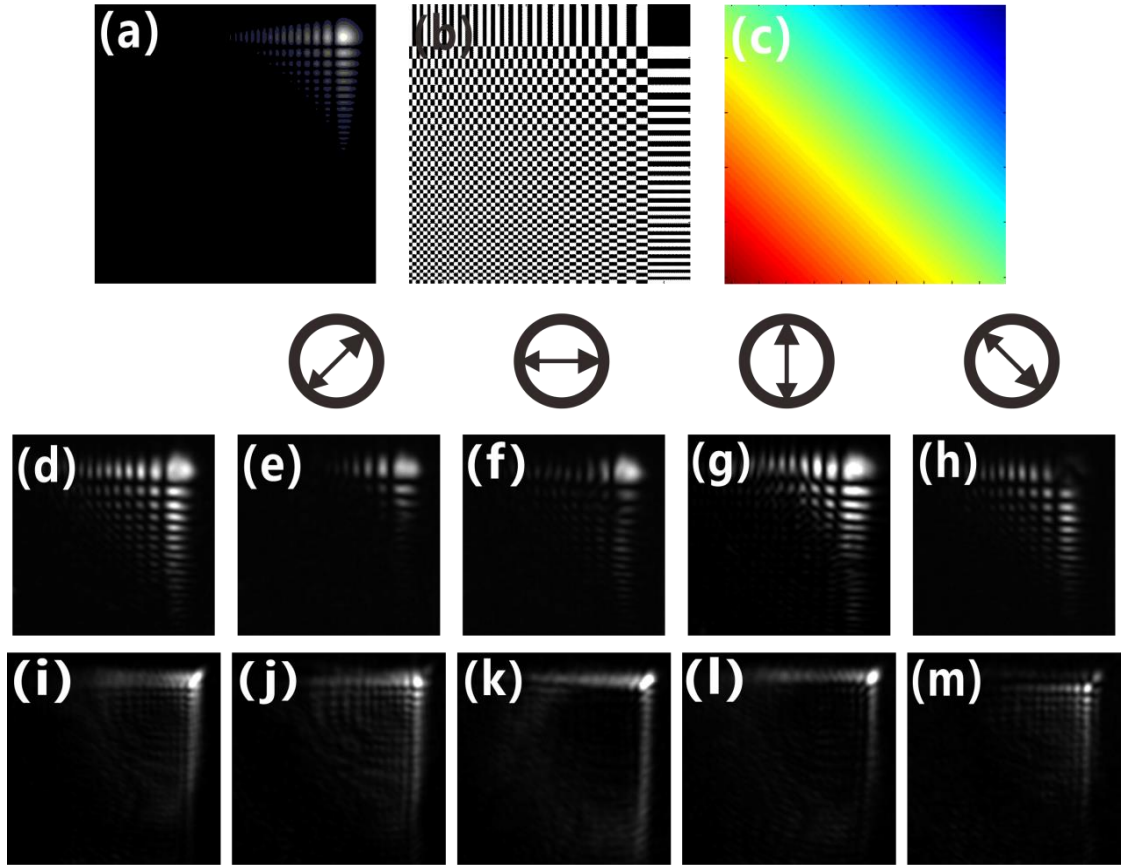

Fig. 1

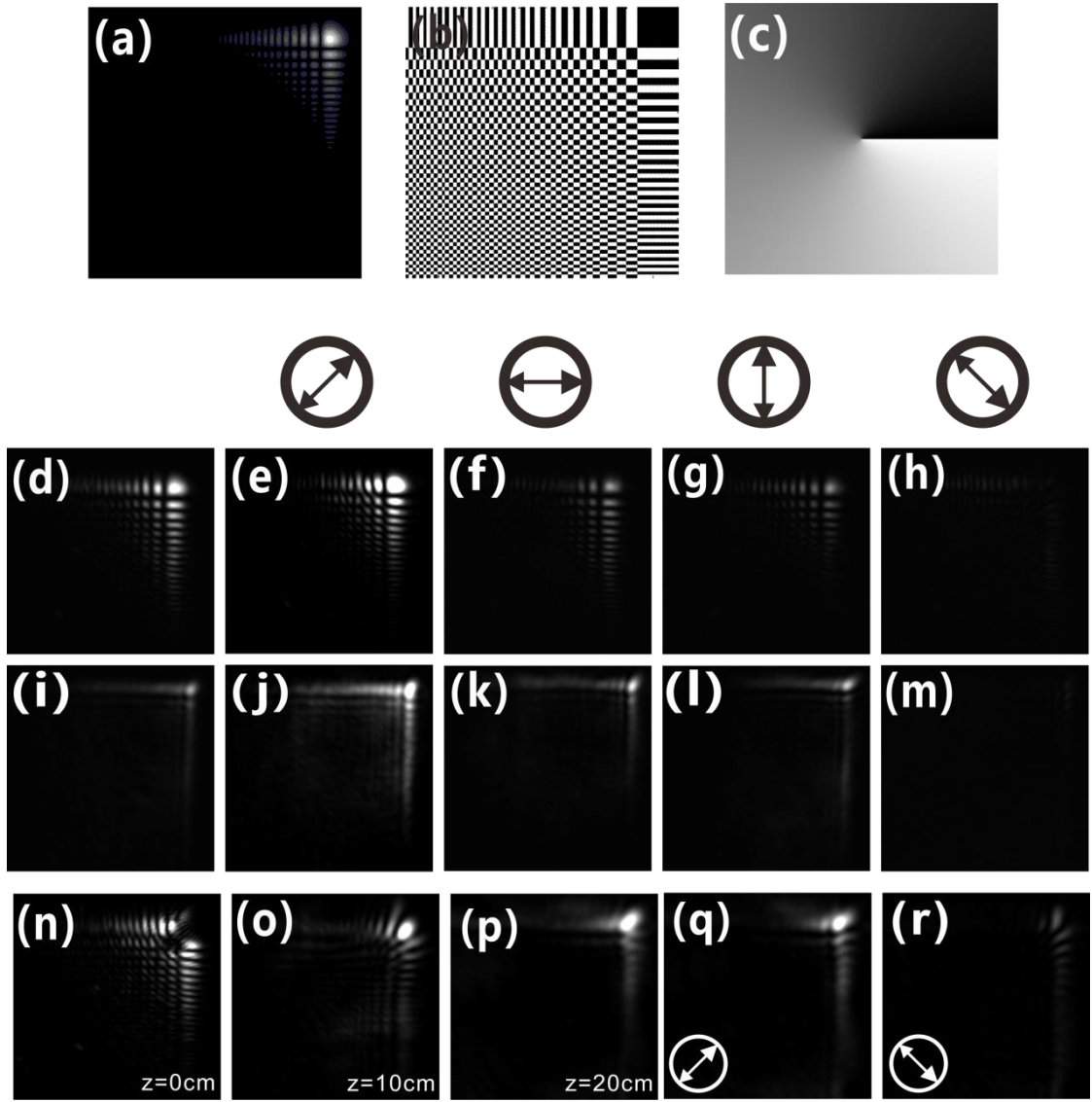

Fig. 2

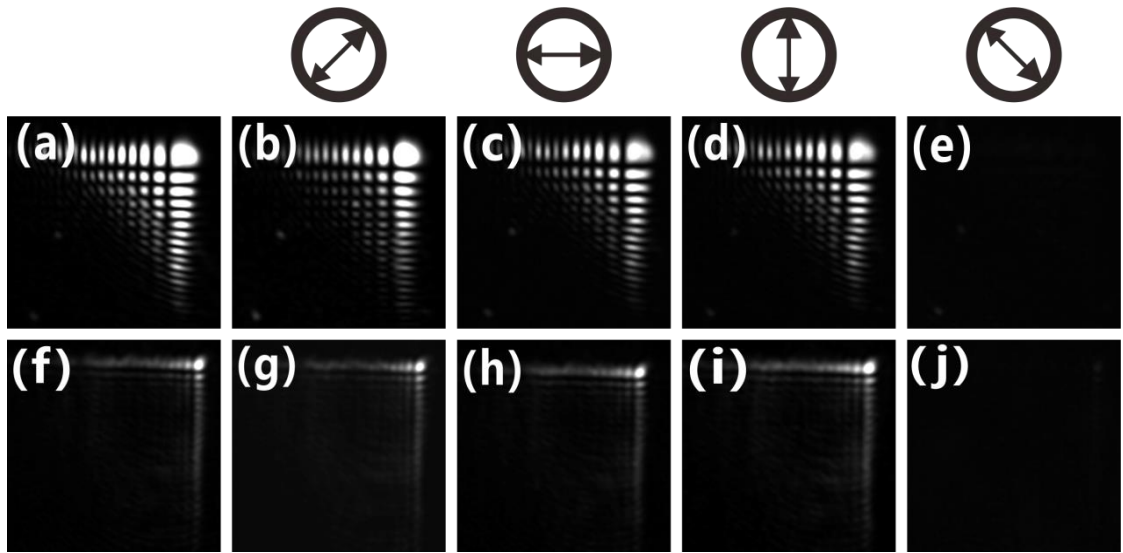

Fig. 3

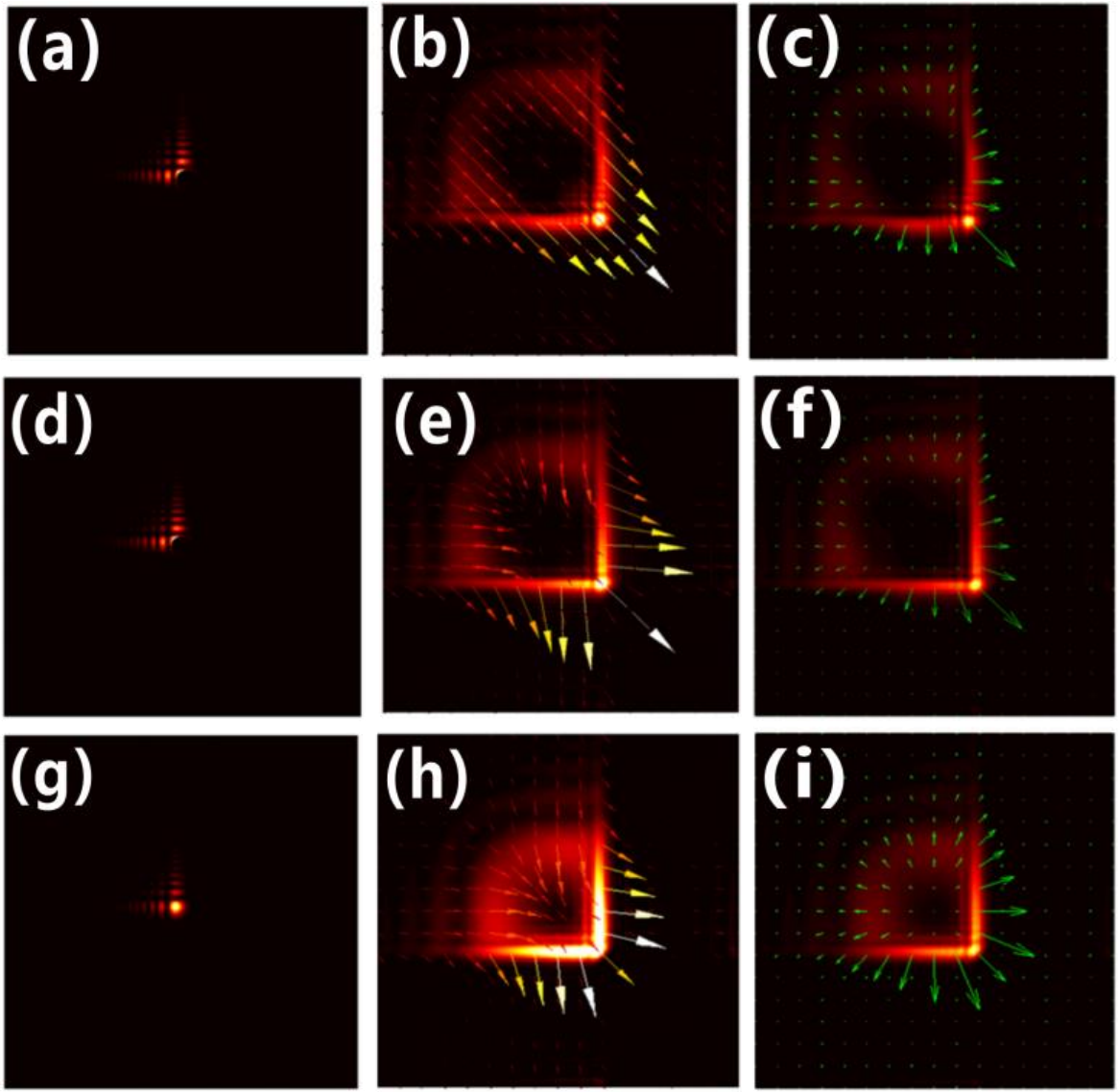

Fig. 4

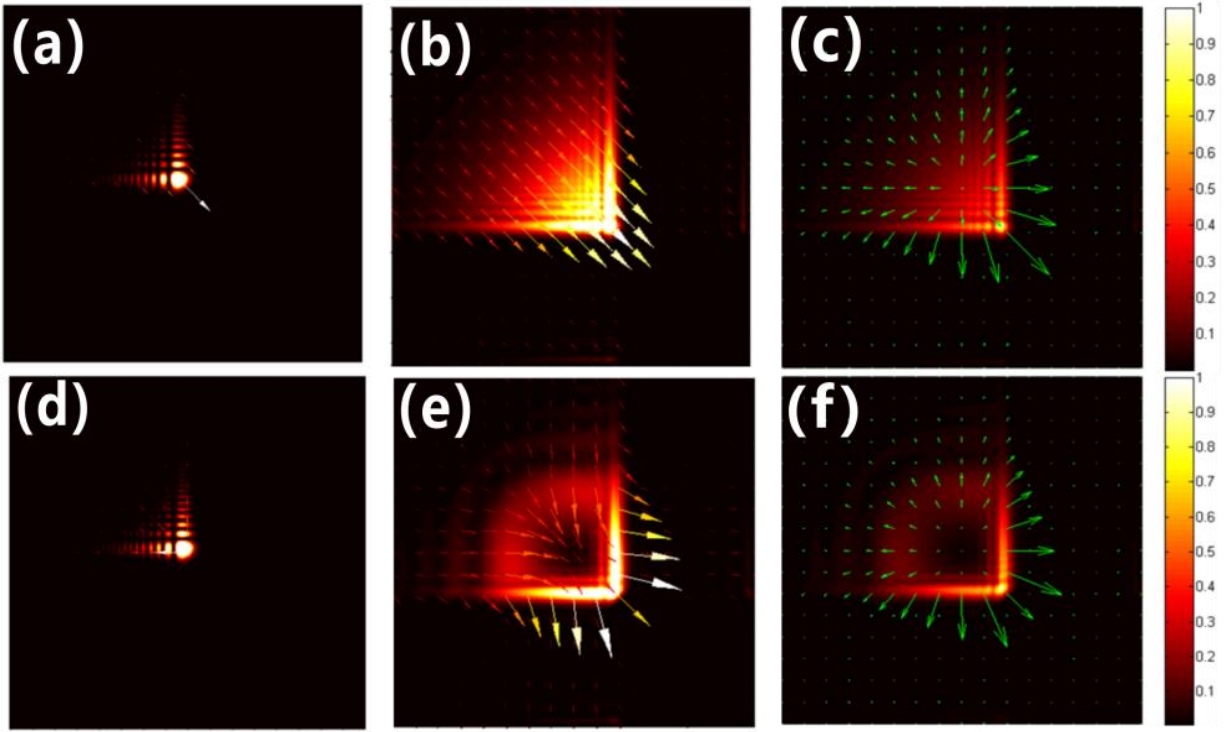

Fig. 5